# Desenvolvimento de modelo para predição de cotações de ação baseada em análise de sentimentos de *tweets*

Mario Akita
*Instituto Federal de São Paulo*
Campinas, Brasil
mario.akita@aluno.ifsp.edu.br

Prof. Me. Everton Josue da Silva
*Instituto Federal de São Paulo*
Campinas, Brasil
everton.silva@ifsp.edu.br

*Abstract*—O treinamento de modelos de aprendizado de máquina para predição de cotações de ações tem sido um assunto cada vez mais abordado à medida que o avanço tecnológico possibilitou o envio automatizado e instantâneo de ordens de compra e venda desses ativos. Enquanto a grande maioria das abordagens nesta disciplina consiste em treinar modelos de Redes Neurais com base somente na cotação histórica dos ativos, neste trabalho utilizamos a plataforma *iFeel* 2.0 para extrair 19 indicadores de sentimentos de postagens da plataforma de *microblogs Tweeter* relacionadas à empresa Petrobras e treinamos modelos XGBoost para prever a cotação das ações desta empresa. Posteriormente, simulamos o desempenho deste modelo e comparamos à média de outros 100 aleatórios para determinar que houve ganho médio de R$88,82 (brutos) no período ao utilizar o modelo treinado, quando comparado ao rendimento médio dos outros cem modelos aleatórios.

## I. INTRODUÇÃO

Desde a digitalização das operações de compra e venda de ativos financeiros intensificada nos anos 1990s, as operações de negociação de ações vêm sendo alvo de intenso estudo com o objetivo de gerar algoritmos que sejam capazes de trazer retorno financeiro aos investidores de maneira automatizada. Recentemente, dada a evolução do poder computacional que viabilizou modelos cada vez mais complexos de negociação, a negociação automática de ações através de algoritmos já representava um montante de pelo menos 50% de todas as negociações de ações nas bolsas de valores dos Estados Unidos no ano de 2012 [1].

Tradicionalmente, os modelos para predição de preços de ação são construídos com base em estatísticas sobre preços, volumes de negociação, médias móveis, dentre outras informações estatísticas e contábeis do ativo financeiro em questão, sendo considerados, portanto, uma evolução da escola de análise técnica de ações [2]. A recente disponibilização de grandes volumes de dados e incrementos no poder de processamento criou um ambiente propício para o desenvolvimento de novos algoritmos mais complexos[3] como diferentes redes neurais ou combinação de vários algoritmos clássicos que possibilitaram a utilização de outros aspectos não estatísticos como indicadores de sentimento em notícias [4], ou tweets [5] e [6].

Neste trabalho, utilizaremos as mais recentes técnicas de Processamento de Linguagem Natural (NLP) para extrair 19 indicadores de sentimentos através de modelos já existentes na literatura que são, posteriormente, utilizados como *features* juntamente com estatísticas de preços e volumes de negociação para treinamento de modelo computacional XGBoost com o objetivo de prever cotações futuras das ações preferenciais da empresa Petróleo Brasileiro S.A. – Petrobras (PETR4).

## II. OBJETIVOS

Neste projeto, desenvolvemos modelos computacionais para prever variações nos preços da ação preferencial da Petrobras (PETR4). O objetivo do projeto é desenvolver modelos que apresentem melhor desempenho quando comparados a modelos aleatórios e que consigam melhores ganhos financeiros do que a realização de operações ao acaso dentro do intervalo de tempo reservado para testes.

## III. BASES TEÓRICAS

*A. iFeel 2.0*

Aplicação Web implementada por Araujo *et al*. [7] para simplificar a implementação e o uso de múltiplos métodos de análise de sentimentos no nível de sentenças. São suportados 19 modelos de análise de sentimentos no total brevemente explicados a seguir:

*1) Emoticons:* Proposto por Gonçalves *et al*. [8] atribui uma pontuação de sentimentos baseado nos *emoticons* utilizados dentro da frase.

*2) Happiness Index:* Proposto por Dodds *et al*. [9], consiste em uma escala de 1 a 9 em que frases são classificadas de acordo com um uso de 1034 palavras e suas escalas na Affective Norms for English Words (ANEW) [10].

*3) SentiWordNet:* É uma ferramenta proposta por Esuli *et al*. [11] comumente utilizada na classificação de opiniões baseada em um dicionário léxico chamado WordNet que considera palavras em língua inglesa. O modelo agrupa palavras em conjuntos chamados de *synsets*. Em seguida, de acordo com as palavras e intensificadores deste conjunto, calcula uma pontuação para considerar o sentimento positivo ou negativo de cada *synset*. Ao final, todos os *synsets* são ponderados para calcular a polaridade global da sentença.

*4) Senticnet:* Proposto por Camnria *et al*. [12], é um modelo que utiliza técnicas de NLP baseado em inteligência artificial. Contém 14 mil conceitos que são utilizados para calcular a polaridade da sentença e foi inicialmente utilizado para avaliar os comentários de pacientes do National Heath System (NHS) na Inglaterra.

*5) PANAS-t:* É uma escala psicométrica proposta por Gonçalves *et al.* [13] para detectar humor baseado no

método PANAS (*Positive Affect Negative Affect Scale*) que analisa o texto diante de nove categorias de humor. Na implementação do *iFeel*, para gerar a escala de polaridade global a nível de sentença, foram considerados 4 humores como polaridade positiva, quatro como polaridade negativa e um neutro.

*6) Sentistrength:* Este modelo [14] propõe uma mistura de uma série de métodos de classificação supervisionados ou não (como regressão logística, *Support Vector Machine* (*SVM*), árvores, dentre outros) para avaliar a polaridade do texto utilizando uma extensão do dicionário LIWC.

*7) SASA:* O *SailAil Sentiment Analyzer*[15] é baseado em técnicas de aprendizado de máquinas similares ao *SentiStrength* e foi desenvolvido para analisar postagens do *Twitter*. Na implementação do *iFeel*, foi utilizado um classificador *Naive Bayes* treinado pelos autores do método.

*8) Opinion Lexicon:* Focada originalmente em avaliações de produtos, o modelo *Opinion Lexicon* proposto por Hu *et al.* [16] construiu um dicionário léxico para capturar se as frases utilizadas em avaliações de produtos na internet eram positivas ou negativas

*9) Opinion Finder (MPQA):* Focado em identificar aspectos subjetivos das frases utilizando análise léxica e modelos de aprendizado de máquina. Foi proposto originalmente por Wilson *et al.* [17] e [18].

*10) AFINN:* Construído a partir de postagens da rede do *Twitter* para melhor capturar o linguajar utilizado na plataforma, o dicionário léxico proposto por Nielsen [19] é considerado uma expansão do dicionário ANEW[20] que propunha atribuir pontuações a palavras em inglês de acordo com a emoção a que elas eram associadas.

*11) SO-CAL:* Proposto por Taboada *et al.* [21], cria um dicionário léxico que contém n-gramas associados a uma escala de emoções que varia de -5 a +5 de modo a capturar palavras que intensificam o sentido de outras ou expressões ao invés de classificar palavras isoladamente.

*12) Emoticons Distant Supervision:* Criado a partir de uma ampla base de tweets, o modelo proposto por Hannak *et al.* [22] calcula a polaridade através do cálculo da frequência em que cada componente léxico aparece na frase analisada.

*13) NRC Hashtag:* O modelo proposto por Mohammad [23] pontua frases de acordo com a frequência em que determinadas *hashtags* são utilizadas na frase.

*14) Emolex:* Constrói um dicionário léxico associado a 8 emoções básicas. Utiliza unigramas e bigramas para associar as pontuações e foi proposto e construído colaborativamente por Mohammad *et al.* [24].

*15) SANN*: Criado para recomendação de conteúdos multimídia no *TED Talks*, o modelo [25] utiliza ações do usuário como indicadores para inferir sentimentos dos usuários o os combina com comentários não classificados para gerar um modelo *Sentiment-aware nearest neighbour model* (SANN).

*16) Sentiment140 Lexicon:* Utiliza um dicionário léxico calculado com base no *dataset* utilizado para treinar o método *Sentiment140*. Proposto por Mohammad *et al.* [26], a pontuação de cada expressão contida no dicionário foi calculada levando em conta a utilização de *emoticons* em *tweets* e a frequência de utilização de uma expressão nos *tweets* de determinada classe.

*17) Stanford Recursive Deep Model:* Introduz um modelo *Recursive Neural Tensor Network* (RNTN) que computa cada frase de acordo com a maneira em que seus componentes interagem entre si dentro da frase. Proposta por Socher *et al.* [27], utiliza recursividade para levar em consideração, por exemplo, a posição em que cada palavra ou expressão parece na frase na hora de calcular a polaridade de cada frase.

*18) Umigon:* O modelo proposto por Levsallois [28] utiliza heurísticas para detectar negações, palavras alongadas e avalia *hashtags* com o objetivo de desambiguar frases.

*19) Vader:* Validado por humanos, o método proposto em 2014 [29] foi criado a partir de dados do *Twitter* e outras redes sociais. Utiliza uma abordagem que valoriza a opinião humana uma vez que se utilizou de avaliadores humanos e coleta de opinião de massas em seu desenvolvimento.

*B. XGBoost*

É um método para treinamento de árvores de decisão (*decision trees*) baseado em aumento do gradiente (*gradient boosting*) [30]. Muito utilizado em competições de aprendizado de máquina, possui uma performance comparável ao estado da arte estando presente, por exemplo, em 17 das 29 soluções vencedoras de desafios da plataforma *Kaggle* em 2015 [31].

Foi escolhido para o treinamento dos modelos deste trabalho por apresentar grande versatilidade e bom desempenho na solução de diversos problemas de aprendizado de máquina e pela experiência anterior em utilizá-lo para predição de preços de ações.

IV. METODOLOGIA

Todos os scripts foram desenvolvidos em *python* 3 utilizando notebooks da plataforma *jupyter* nos ambientes *anaconda* 3 instalados em computador pessoal e através da plataforma *Google Colab PRO* (exceto a tradução dos *tweets*, que foi realizada no *Google Spreadsheets*).

Um desenho esquemático do *pipeline* implementado pode ser observado na Fig 1.

*A. Aquisição de cotações históricas de ações*

As cotações para as ações preferenciais da companhia Petróleo Brasileiro AS – Petrobras (PETR4) foram obtidas através da plataforma *Meta Trader 5*, disponibilizada gratuitamente para clientes da *Clear* Corretora (XP Investimentos CCTVM S.A.). Por questões operacionais existe um limite de dados disponíveis na plataforma e, para este estudo, foram obtidos dados referente ao período de 3/12/2018 a 28/11/2022. Os dados obtidos têm granularidade de 5 minutos.

O dataset obtido é composto pelos seguintes campos:

- **Date**: Data do período considerado.
- **Time**: Hora de início do período considerado.
- **Open**: Preço da primeira negociação do período.
- **Hig**h: Maior preço negociado no período.
- **Low**: Menor preço negociado no período.

- **Close**: Preço da última negociação do período.
- **Tickvol**: Quantidade de negócios realizados no período considerado.
- **Vol**: Quantidade de ações negociadas no período.
- **Spread**: Indica se houve ou não *spread* (diferença no preço de compra e venda) nos negócios do período.

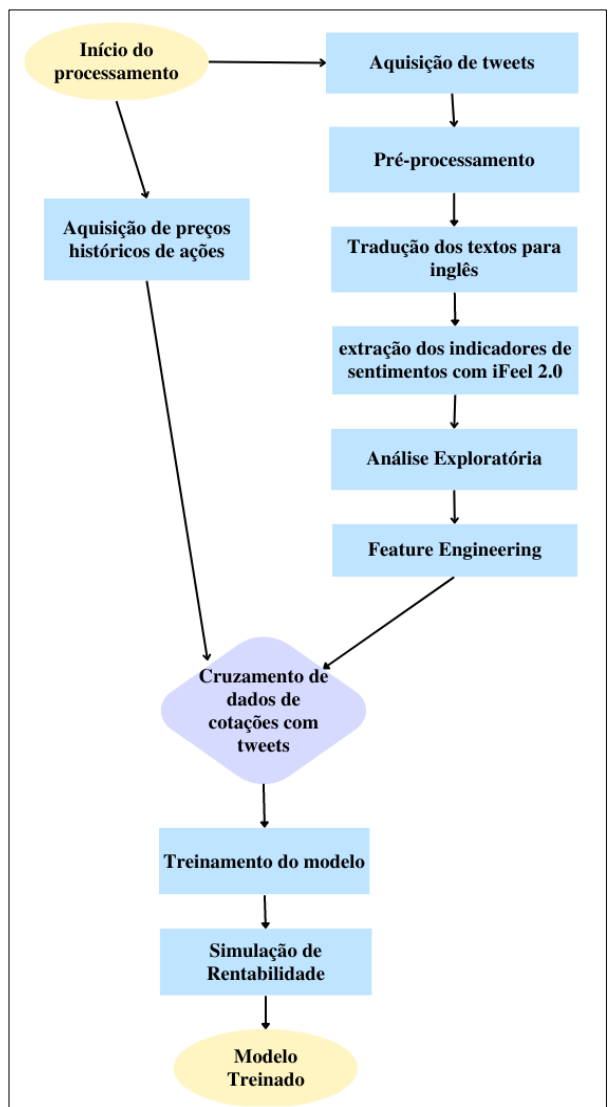

Fig. 1. Representação esquemática do pipeline implementado neste trabalho.

*B. Aquisição de tweets*

A aquisição de *tweets* foi realizada através de script em *Python* desenvolvido pelo autor. Os dados foram obtidos através da API v2 oficial do *Twitter* [33] utilizando a biblioteca *Tweepy*. Foi concedido o status de "pesquisador" para o uso acadêmico da API que possibilitou o incremento da quantidade de *tweets* consultados.

Para construção da base, utilizamos, inicialmente, todos os *tweets* postados entre 13:30 e 19:50 GMT – o que corresponde ao horário habitual de negociação das ações na Bolsa de Valores brasileira – do período de 23/8/2021 a 30/6/2022. Posteriormente, a base foi aumentada para conter os *tweets* criados entre 1/1/2021 e 30/11/2022.

A consulta filtrou apenas tweets que continham as *hashtags* #PETR3 ou #PETR4, além daqueles que continham o nome da empresa "Petrobras". Foram excluídos os *retweets* e aqueles escritos em outra língua que não o português.

Os dados contidos no *dataset* de *tweets* são os seguintes:

- **Created_at**: *timestamp* correspondente à data e hora em que a postagem foi publicada.
- **Text**: o texto da postagem.
- **Like**: quantidade de "curtidas" da postagem.
- **Quote**: quantidade citações da postagem.
- **Reply**: quantidade de respostas da postagem.
- **Retweet**: quantidade de usuários que republicaram a postagem em suas contas.
- **User_followers**: quantidade de seguidores do usuário autor da postagem.
- **User_following**: quantidade de usuários que o autor da postagem segue.
- **User_tweets**: quantidade total de postagens do autor.
- **User_listed**: quantidade de listas criadas por outros usuários que contém o autor da postagem.

Devido ao limite imposto pela API do *Twitter* de retorno de, no máximo, 300 *tweets* por consulta, 180 consultas a cada 15 minutos e uma consulta por segundo, houve a necessidade de dividir as consultas por período de tempo e retardar o processamento das consultas utilizando o comando *sleep()* da linguagem *Python* e limitar manualmente as consultas para que apenas um mês fosse consultado a cada execução do *script*.

*C. Pré-processamento dos tweets*

Após a consulta dos *tweets* postados nos períodos pretendidos, foram efetuadas operações para pré-processar os dados obtidos e retirar alguns elementos que não seriam utilizados.

A primeira medida tomada foi a filtragem dos *links* presentes nas postagens. Por se tratarem apenas de endereços de *sites*, foram filtrados pois não influenciam no cálculo de polaridade dos *tweets*. Outro elemento eliminado das bases foi a menção a outros usuários. Menções são caracterizadas pelo caractere "@" seguido do nome de algum usuário da rede e servem como uma espécie de *link* para marcar outros usuários em postagens que podem interessar a eles. Foram filtrados pois não representam informação relevante para o cálculo da polaridade dos *tweets*.

Foram excluídos, ainda, símbolos não alfabéticos, *emojis* e pontuações. A ideia por trás dessa eliminação foi tentar facilitar obtenção das métricas de polaridade dos *tweets* em um contexto de desenvolvimento de modelos de NLP próprios específicos para este trabalho (como originalmente pensado). Entretanto, o simples desenvolvimento de um modelo de linguagem é uma tarefa complexa que exige quantidades massivas de dados para o treinamento consequentemente a ideia do desenvolvimento acabou substituída. Porém, uma vez que o dataset já havida sido processado quando da mudança de curso, os *emojis* e símbolos removidos acabaram comprometendo o cálculo da polaridade em alguns dos modelos utilizados.

Outro processamento efetuado foram a separação de data e hora e conversão para o fuso horário brasileiro (GMT-3). Necessária para a correta atribuição dos preços e estatísticas obtidos no item anterior aos tweets, os procedimentos visam a padronização de datas e horas para evitar erros e simplificar visualização do pipeline construído. Outra operação efetuada foi a eliminação de duplicatas, que nada mais é do que um procedimento de saneamento de *datasets* corriqueiro, mas importante para evitar contaminação dos dados quando divididos em treinamento e teste. Sua falta poderia prejudicar o cálculo das estatísticas de cada período e, consequentemente, o treinamento do modelo como um todo.

Finalmente, foram eliminados os *tweets* com pouca informação. Postagens com 2 palavras ou menos assim como aquelas com menos de 20 caracteres foram eliminadas pela alta chance de não conterem informações relevantes.

*D. Tradução do texto dos tweets*

Devido à carência de modelos treinados especificamente para o idioma português brasileiro, o sistema *iFeel* 2.0 conta com uma ferramenta de tradução embutida. Apesar disso, o módulo de tradução apresentava erro de conexão com a API e estava indisponível nos períodos de testes. Diante deste cenário, a tradução dos textos dos *tweets* que foram extraídos foi alcançada através da utilização das funções integradas ao *Google Spreadsheets*.

Esta função utiliza internamente o *Gooogle* Tradutor para prover as traduções e foi necessária uma vez que todos os modelos utilizados pelo iFeel 2.0 utilizam-se de dicionários léxicos em inglês ou foram treinados utilizando-se bases textuais naquele idioma.

*E. Extração dos indicadores de sentimentos*

A extração das polaridades de sentimentos foi executada utilizando o programa *iFeel* 2.0. Disponível para *download* em uma imagem *Docker*, o programa foi executado mês a mês em três ambientes: um computador portátil pessoal, um computador pessoal de mesa e um servidor em nuvem.

O processamento se mostrou desafiador, já que congelamentos e travamentos da máquina virtual Java utilizada no projeto ocorreram frequentemente e necessitaram de constante intervenção manual. Ao final deste processo, os *datasets* mensais foram combinados para formar um grande conjunto de dados de 323.460 amostras e 39 atributos que abrangem conteúdo publicado em um intervalo de tempo de 22 meses.

*F. Análise exploratória*

Após a extração de dados, foi efetuada uma breve análise exploratória de dados para melhor compreensão do conjunto de dados a ser trabalhado. A seguir, alguns dos principais achados são discutidos:

*1) Balanceamento de classes:* Há um leve desbalanceamento de amostras. Enquanto são 139.885 as amostras para períodos de alta nas cotações (~56,5% do total), as amostras referentes a períodos baixa somam 107.718 amostras (~43,5% do total).

*2) Distribuição dos atributos de sentimentos:* De maneira geral, a grande maioria dos tweets foram classificados como neutros com as demais postagens se distribuindo de maneira razoavelmente equilibrada dentre as duas polaridades. Há exceções como o método *PANAS-t* que classificou quase todas as amostras como neutras, *EMOLEX* e *OPINION LEXICON* que classificaram muito mais amostras como negativas. Outros modelos, como *EMOTICONS* e *EMOTICONS*-DS foram, obviamente, prejudicados pela limpeza de *emoticons* realizada no pré-processamento. A fig 2 traz a distribuição de algumas das *features* extraídas.

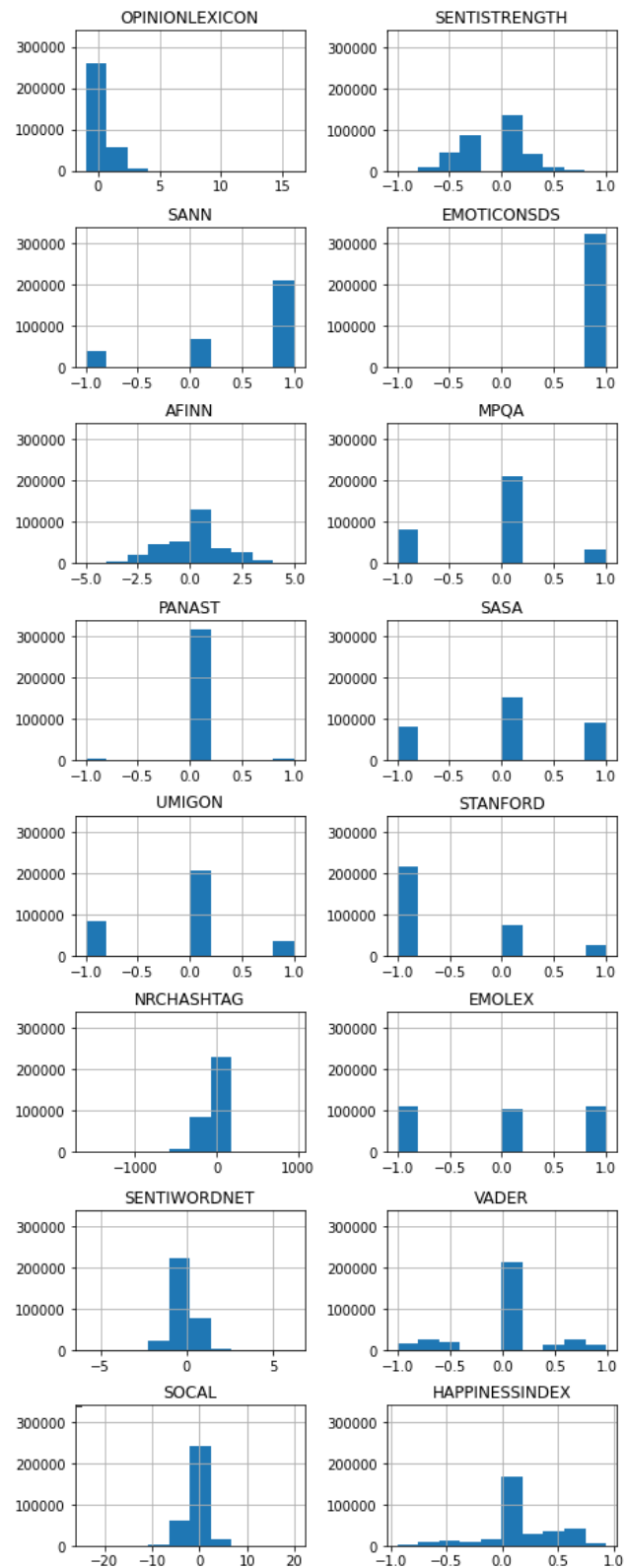

Fig. 2. Distribuição de pontuação das amostras de alguns modelos.

*3) Estudo de correlações:* A variável-alvo não apresentou correlação significativa com nenhum dos outros atributos sendo que as únicas correlações importantes foram registradas entre os valores de polaridade calculados – o que, de certa forma, é esperado uma vez que muitos modelos podem ser treinados a partir de bases textuais similares.

*G. Feature engineering*

Nesta etapa, com o intuito de aumentar a quantidade de *features* no dataset, foram criados atributos com o intuito de prover outras informações que talvez fossem relevantes durante o treinamento, como a hora de criação e tamanho dos *tweets*. A ideia é aumentar a variabilidade de *features* a serem apresentadas para o treinamento com outros dados que não somente aqueles referentes à análise de sentimentos.

*1) Hora:* número inteiro correspondente à hora em que a postagem foi publicada.

*2) Word_count:* contagem de palavras na publicação. Foi utilizada, também, para uma limpeza de dados que eliminou todas as postagens com menos de 3 palavras.

*3) Text_length:* tamanho total do *tweet*. Pensada para ser utilizada como um indicador de confiabilidade das *features* de polaridade durante o treinamento, parte do princípio de que quanto maior a publicação, maior a quantidade de palavras consideradas no cálculo da polaridade e, por consequência, mais "confiável" é a métrica.

Para formar o *dataset* final, foram utilizadas estatísticas de todos os *tweets* contidos em cada intervalo para que cada intervalo de tempo fosse representado por uma única amostra de dados, e, desta forma, criar uma base balanceada em número de amostras por período e inferir uma sequência entre as amostras de cada período.

Para cada um dos 27 atributos dos tweets e polaridades calculadas, foram utilizadas as seguintes estatísticas:

- Média
- Desvio-padrão
- Mínimo
- Máximo
- Soma
- Variância
- Contagem da quantidade de amostras do período

Além das estatísticas, foram adicionadas as *features* com defasagem destas estatísticas e dos atributos de preços e volumes de negociação. Desta forma, criou-se uma relação temporal entre os dados, o que é importante pois durante o treinamento o modelo será exposto a dados do tempo atual assim como dados dos tempos anteriores.

Foram utilizadas as seguintes janelas de tempo para adição dos atributos com atraso:

- Atraso de 5 minutos
- Atraso de 10 minutos
- Atraso de 15 minutos
- Atraso de 20 minutos

*H. Cruzamento de dados de cotações com tweets*

Devido às mudanças de estratégias ocorridas durante o desenvolvimento do projeto, este passo foi realizado em diferentes pontos do *pipeline* de dados executado. Apesar disto, por se tratar de um cruzamento de dois conjuntos de dados, ocorre apenas a adição de novas *features* para cada amostra, e sua presença ou ausência não interfere nos processamentos das etapas anteriores.

Este passo busca atribuir os preços e estatísticas de negociações da ação com os *tweets* que foram publicados naquele momento, portanto, o horário de postagem de cada *tweet* é arredondado para o múltiplo de 5 imediatamente inferior e as cotações desse tempo são atribuídas à postagem.

Adicionalmente, o conjunto de dados também recebe seu atributo-alvo. Como o objetivo do trabalho é prever a cotação da ação no próximo período de tempo (5 minutos), atribuímos o preço de fechamento em t + 5min a cada amostra para utilizarmos como alvo durante o treinamento.

*I. Treinamento dos modelos*

Foi treinado um modelo XGBoost para realizar todas as previsões do dia utilizando os seguintes parâmetros de treinamento:

- ETA: 0.01
- N_ESTIMATORS: 300
- RANDOM_STATE: 4321
- SCALE_POS_WEIGHT: 0,6
- MAX_DEPTH: 5
- OBJECTIVE: "binary_logistic"

*1) Separação entre conjunto de teste e validação*

Antes de definir quais os dados serão considerados na hora da divisão, é necessário definir a quantidade de dias a serem incluídos em cada grupo. Por padrão, foram utilizados os seguintes valores:

- Dias de treinamento: 200
- Dias de validação: 1
- Dias de teste: 1

*2) Treinamento do modelo de treinamento*

Desta forma, por exemplo, para treinar o primeiro modelo de validação, no *dataset* criado, utilizaríamos, como dados de treinamento, todas as amostras compreendidas entre 4/1/2021 e 26/10/2021. As amostras do dia seguinte (27/10/2021) são utilizadas para validação de forma a otimizar o modelo para este período. Note-se, entretanto, que os dados a serem de fato classificados são os do dia 28/10/2021 – que não são utilizados neste primeiro treinamento.

*3) Treinamento do modelo de validação*

Para gerar o modelo que fará a previsão de dados, retreina-se o melhor modelo obtido no passo anterior com dados dos dias de treinamento adicionados dos dados do dia de validação. O desempenho somente é aferido utilizando os dados do dia reservado para testes.

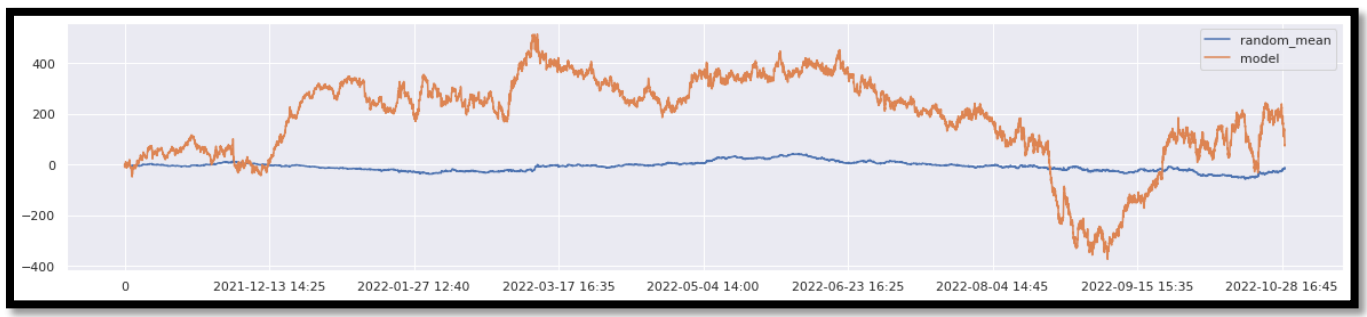

Fig. 3. Gráfico mostrando a rentabilidade no período de testes.

*J. Simulação de desempenho*

Foi implementado um calculador de desempenho para verificar qual seria o resultado financeiro ao se aplicar o algoritmo de predição treinado simulando operações de compra e venda da ação, de acordo com o resultado previsto pelo modelo. Foi utilizado um conjunto de 100 modelos aleatórios como *baseline* para entender se o modelo treinado realmente produz melhores resultados quando comparado à média dos modelos aleatórios.

Para tanto, foi utilizada a própria tabela de cotações e as seguintes regras:

- Caso o valor predito pelo modelo (ou o valor escolhido aleatoriamente, para os modelos aleatórios) seja 1, é simulada uma compra da ação no preço de abertura seguida de uma venda ao final do período.
- Caso contrário faz-se uma venda descoberta do papel, que consiste na venda do papel pelo preço de abertura seguida de uma compra ao final do período.
- Calcula-se a rentabilidade diária de cada modelo de acordo com a tabela a seguir:

TABLE I. CÁLCULO DA RENTABILIDADE DIÁRIA

| Previsão | Fórmula |
|---|---|
| 0 | $Rentabilidade = QTDE * (P_{CLOSE} - P_{OPEN})$ |
| 1 | $Rentabilidade = QTDE * (P_{OPEN} - P_{CLOSE})$ |

a. P refere-se ao preço do ativo na abertura (OPEN) e fechamento (CLOSE)

Para todos os efeitos de cálculo de rentabilidade, foram desconsideradas eventuais taxas de corretagem, emolumentos, impostos, juros e quaisquer outras cobranças. A rentabilidade final foi calculada sempre considerando a negociação de 1 lote de ações (100 ações).

*K. Apresentação e Análise de resultados*

O modelo treinado apresentou excesso de rentabilidade bruta de R$77,00 frente à média de -R$11,82 dos modelos aleatórios durante o período de testes entre 27/10/2021 e 28/10/2022. Ou seja, no período considerado houve um ganho médio de R$ 88,82 ao escolher o modelo treinado em detrimento de um modelo aleatório hipotético aqui representado pela média de 100 modelos aleatórios.

Um gráfico com a rentabilidade bruta obtida em cada período pode ser observado na figura 3. Nele, podemos observar que na maioria do período estudado, o modelo treinado esteve acima do modelo aleatório médio. Isto evidencia uma relativa consistência do modelo treinado em entregar os resultados melhores que o aleatório.

A seguir apresentamos as métricas *Macro* do modelo treinado durante o período de validação:

- Precisão: 0,51
- *Recall*: 0,52
- F-1: 0,40
- AUC: 0,5153
- *Logloss*: 0,6923

*L. Breve discussão de resultados*

Apesar das métricas *macro* não serem consideradas boas, o modelo proposto conseguiu obter algum resultado positivo frente a um conjunto de modelos operando aleatoriamente, o que demonstra uma certa consistência do modelo em entregar resultados diante do intervalo de tempo relativamente grande (cerca de 1 ano).

Quando consideramos os períodos em que o modelo ganhou ou perdeu da média dos modelos aleatórios, podemos observar que o modelo treinado ganhou em 215 períodos e perdeu em 37 deles.

Houve um lucro bruto calculado de R$ 77,00 durante o período reservado para testes, o que representa um ganho médio de R$0,31 por operação, considerando a compre e venda de um lote de ações (100 ações). A média dos cem modelos que operaram aleatoriamente, gerou um ganho total no período de R$ -11,82, o que equivale um ganho médio de R$ -0,05 por operação.

Algumas conjecturas de fatores que podem ter interferido com o desempenho do modelo são:

- Dados talvez não tenham correlação suficiente com a cotação de ações específicas: A plataforma *Tweeter* em geral, tende a ser cenário de disputas e reclamações, o que nos levaria a ter muito mais conteúdo de opiniões negativas do que positivas e isto nem sempre condiz com as cotações de ações. Este fenômeno pode ter acontecido já vez que apesar do *dataset* de preços ser levemente desbalanceado para o campo positivo, as polaridades são levemente mais negativas que positivas. Além disso, situações como o aumento do preço de combustíveis, por exemplo, podem aumentar a quantidade de críticas nas redes sociais enquanto as ações sobem, pelo possível aumento no faturamento e, posteriormente, nos lucros da empresa.
- Período eleitoral: Parte considerável dos dados de testes se deu em período eleitoral, o que pode ter

gerado um descolamento entre os dados coletados e as cotações já que a Petrobrás é uma empresa estatal. Este fato é melhor observado no período em que o desempenho do modelo ficou abaixo da média dos modelos aleatórios já na parte final do gráfico da fig. 3. O período (agosto/setembro de 2022) coincide exatamente com o início das campanhas eleitorais para as eleições gerais daquele ano.

- Tradução: A tradução automática dos tweets de português para a língua inglesa pode ter provocado a perda de conteúdo semântico e prejudicado o cômputo das polaridades, uma vez que expressões comumente utilizadas no Brasil podem ter carga sentimental maior ou menor do que suas equivalentes em inglês.

## V. Conclusão e trabalhos futuros

Considerando toda a extensão do trabalho desenvolvido até aqui, que contemplou desde a aquisição de dados passando pela extração de características, amplo estudo sobre os dados adquiridos até o treinamento e mensuração de resultados, podemos afirmar que foi um trabalho desenvolvido de ponta-a-ponta utilizando apenas recursos disponíveis na internet gratuitamente (exceto *Google Colab PRO*).

Neste sentido, o objetivo de desenvolver um modelo que pudesse superar o aleatório para negociação de ações baseada em indicadores de sentimentos de *tweets* foi considerado alcançado.

Há, entretanto, uma vasta quantidade de pontos em que futuros trabalhos e abordagens alternativas poderão corrigir ou melhorar as falhas cometidas na abordagem aqui descrita e, consequentemente, melhorar os resultados. A seguir, algumas sugestões para futuras continuações deste trabalho.

### A. Expansão do intervalo de tempo analisado

Inicialmente pensado para cobrir uma granularidade maior de dados (intervalo de 1 minuto), encontrar dados disponíveis gratuitamente ou com preço muito baixo para este intervalo de tempo mostrou-se uma tarefa infrutífera. Entretanto, ao migrarmos para o intervalo de 5 minutos, pudemos achar com facilidade dados até o ano de 2018. Uma das ideias é utilizar essa base maior para adquirir mais dados de períodos não considerados neste trabalho e, eventualmente, gerar modelos melhores.

### B. Expansão para ações de outras empresas

Neste trabalho focamos nas ações da empresa Petrobras por ser a ação mais líquida e, por ampla margem, a empresa negociada mais comentada da rede social – principalmente após os sucessivos reajustes de preços ocorridos em 2021 e 2022. Entretanto, a mesma estratégia poderia ser implementada para outras empresas (ou mesmo um conjunto de empresas) em novos trabalhos, aproximando-se da abordagem de publicações que utilizaram índices de ações[5].

### C. Utilização de outras estratégias de modelagem

Neste trabalho focamos na utilização do *XGBoost* como algoritmo de preferência para treinarmos o nosso modelo, porém, há outros algoritmos que também poderiam ser implementados, como por exemplo:

*1) LSTM:* Muito utilizado para predição de dados que possuam um componente temporal entre as amostras[34], poderia ser utilizado para predição de cotações de ações;

*2) SOFNN:* Rede neural que utiliza lógica *Fuzzy* já foi utilizada para prever índices de ações através de dados obtidos com análise de sentimentos de postagens do *twitter* [5] e poderia ser aplicado na resolução deste problema.

### D. Utilização de outras fontes de dados:

Outras redes sociais ou mesmo *feeds* de jornais e notícias poderiam ser utilizados para adicionar confiabilidade aos dados.

### E. Utilização de outra função para medir o desempenho do algoritmo no treinamento

Como pudemos observar na apresentação dos resultados, nem sempre períodos de maior acerto ou mesmo maior AUC apresentam os melhores resultados. Por padrão, a biblioteca *XGBoot* vem com duas funções para auferir o desempenho durante o treinamento: *AUC* e *LogLoss*. Talvez fosse necessária a adição de algum componente de peso a estas funções ou desenvolvimento de função baseada na rentabilidade para melhorar as métricas de rentabilidade dos modelos gerados.